\documentclass[%
 reprint,
 amsmath,amssymb,
 aps,pre
]{revtex4-1}

\usepackage{graphicx}
\usepackage{dcolumn}
\usepackage{bm}
\usepackage{amsmath}
\usepackage{hyperref}

\begin{document}

\preprint{APS/123-QED}

\title{How to Simulate Patchy Particles}

\author{Lorenzo Rovigatti}
\email{lorenzo.rovigatti@gmail.com}
\affiliation{CNR-ISC, Uos Sapienza, Piazzale A. Moro 2, 00185 Roma, Italy}
\affiliation{Dipartimento di Fisica, {\em Sapienza} Universit\`a di Roma, Piazzale A. Moro 2, 00185 Roma, Italy}
\author{John Russo}%
\affiliation{School of Mathematics, University of Bristol, Bristol BS8 1TW, United Kingdom}
\email{john.russo@bristol.ac.uk}
\author{Flavio Romano}%
\affiliation{Dipartimento di Scienze Molecolari e Nanosistemi, Universit\`a Ca' Foscari di Venezia, Via Torino 155, 30172 Venezia Mestre, Italy}
\email{flavio.romano@unive.it}

\date{\today}

\begin{abstract}
\emph{Patchy particles} is the name given to a large class of systems of mesoscopic particles characterized by a repulsive core and a discrete number of short-range and highly directional interaction sites.
Numerical simulations have contributed significantly to our understanding of the behaviour of patchy particles, but, although simple in principle, advanced simulation techniques are often required to sample the low temperatures and long time scales associated with their self-assembly behaviour.
In this work we review the most popular simulation techniques that have been used to study patchy particles, with a special focus on Monte Carlo methods. We cover many of the tools required to simulate patchy systems, from interaction potentials to biased moves, cluster moves, and free energy methods. The review is complemented by an educationally-oriented Monte Carlo computer code that implements all the techniques described in the text to simulate a well-known tetrahedral patchy particle model.
\end{abstract}

\maketitle

\section{Introduction}

The expression ``patchy particle'' has become more and more inclusive over the course of the years~\cite{patchy_review}. The most basic definition, ``a colloid with attractive spots decorating its surface'' is now outdated as patchy particles, or concepts developed in the field of patchy particles, are being used to model proteins~\cite{Coluzza2014,McManus2016,squid_science}, viral capsids~\cite{Rapaport2008}, hard faceted bodies~\cite{Millan2015}, double-stranded DNA~\cite{demichele_dna_patchy} and even atoms and molecules~\cite{Whitelam2015,Sciortino2008}.

On the experimental side, tremendous progress is currently being made in the realization of such systems~\cite{Wang2012colloids,yi2013recent,liu2016diamond,striolo2016janus,zhang2017janus,diaz2018dna}, and patchy particles have already been adopted by theorists as an ideal model to study the self-assembly properties of a variety of soft-matter systems~\cite{bianchi2011patchy,tavares2015generalization,lisbona_dynamics,teixeira2017phase,patchy_review}. The number of examples of the impressive agreement between theory and simulations is rapidly increasing~\cite{chen2011directed,iwashita2013stable,biffi_dna,reentrant_gels,liu2017directionally}, establishing patchy particles as one of the most active and successful ideas in nanotechnology.

Molecular simulations played an important role in the development of the field, providing early predictions for many interesting new phenomena~\cite{kern_frenkel,Zhang2005,bianchi2006phase,doye_patchy,doye_diamond,bianchi:2011}, which sparked the interest of the experimental community~\cite{Pawar2009,Wang2012colloids,Kraft2011,vanOostrum2015,Tigges2015}. Since the early days, many of the numerical and theoretical predictions, such as the shrinking of the gas--liquid phase separation region~\cite{bianchi2006phase,biffi_dna}, the existence of reentrant gas-liquid phase separations and reentrant gels~\cite{roldan2013gelling,reentrant_gels,reentrant_spinodal}, or the lack of crystallisation in highly flexible patchy systems~\cite{frank_nat_phys,Rovigatti2014a,biffi_soft_matter}, have been observed in experiment. Even though it is becoming more and more common to consider patchy particles of increasing complexity~\cite{patchy_review}, the majority of the theoretical breakthroughs have been obtained with toy models. Such simple models usually feature hard spherical particles and short-range patch-patch attractions.
Notwithstanding the raise in popularity of systems composed of or inspired by patchy models, to the best of our knowledge a comprehensive resource to help choose and implement the right state-of-the-art algorithm to tackle the investigation of these systems is still lacking. This Review aims at filling such a gap.

In the following we give an overview of the more commonly used models and simulation methods that have been developed or adapted to patchy particles. We put special emphasis on Monte Carlo (MC) simulations, but we also mention models and algorithms pertaining to molecular dynamics (MD) simulations. In order to help the interested reader, we have implemented the most important MC algorithms described below in an open-source code, freely accessible on the web (\href{http://dx.doi.org/10.5281/zenodo.1153959}{http://dx.doi.org/10.5281/zenodo.1153959})~\cite{pp_code}. This \textit{PatchyParticles} code (from here on referred to as the \textit{PP code}) has been developed with the idea of providing a simple and clear implementation rather than a fast and optimised one, and hence should be considered as an educational tool that complements the Review rather than a production-ready code.

\section{Patch-patch interaction potentials}\label{sec:potentials}

\begin{figure*}[t!]
\includegraphics[width=0.8\textwidth]{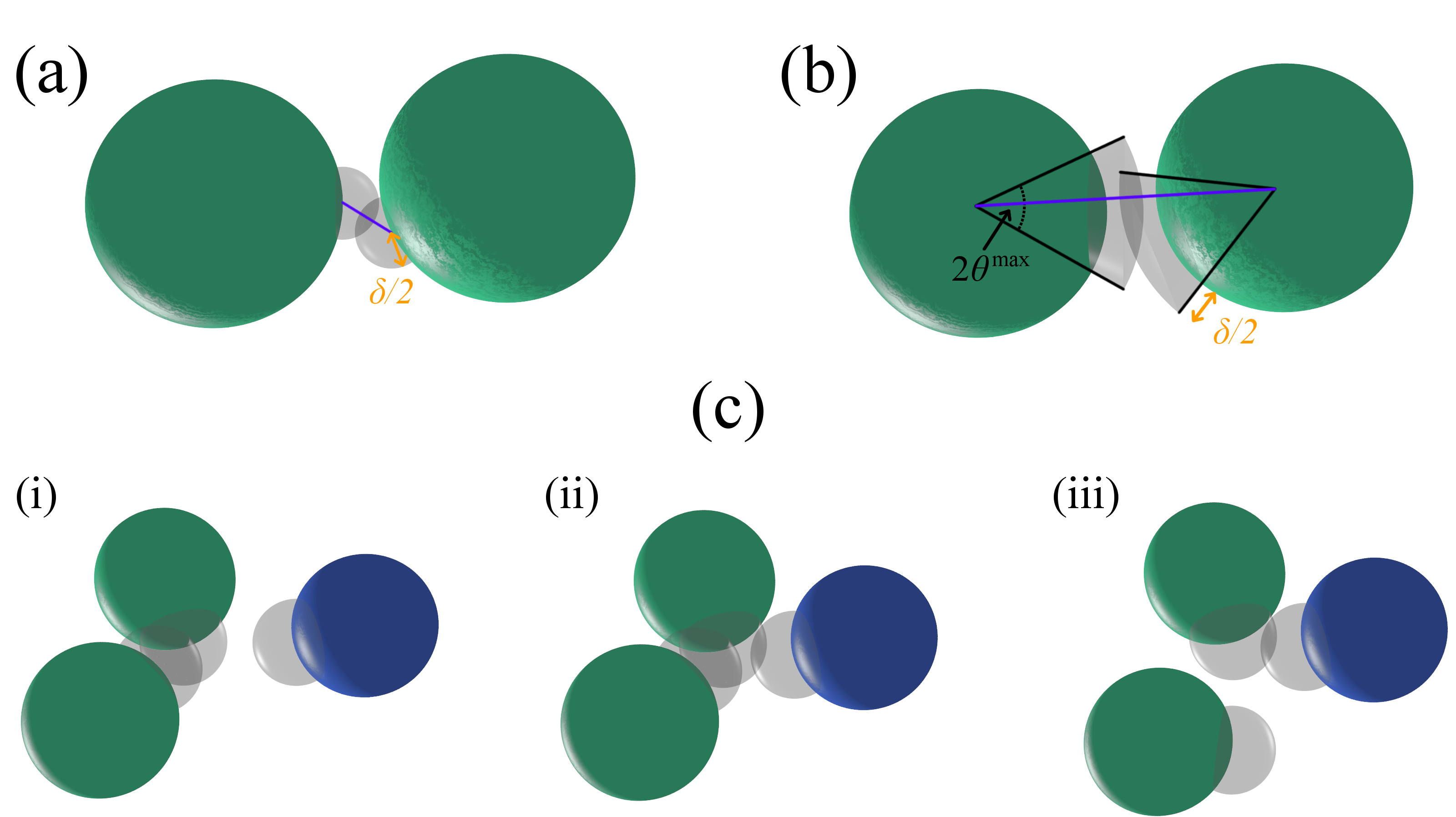}
\caption{\label{fig:snapshots}A cartoon showing particles decorated with single patches, depicted in light grey. (a) In the sticky-spot model two patches interact only if their mutual distance (in violet) is smaller than two times the patch radius, $\delta/2$ (in orange). (b) In the Kern-Frenkel model patches are modelled as spherical cones of angular width $2 \theta^{\rm max}$ and length $(\sigma + \delta)/2$, where $\sigma$ is the particle diameter. Any two patches interact if they intersect with each other and if the vector connecting the centres of mass of the two particles (in violet) pass through both of them. (c) Thanks to the bond-swapping algorithm (see text and Ref.~\cite{Sciortino2017}), a free patch (here attached to the blue particle) can establish a bond with an already bonded patch whose original partner is free to move away at the end of the process. The attractive contribution due to the additional patch-patch bond present in (ii) is (partially or totally) counterbalanced by a three-body repulsion. Indeed, the energetic cost of the bond-swapping process, that is, its activation energy, is controlled by a parameter of the model and can be set to 0 make the reaction temperature-independent.}
\end{figure*}

The inter-particle interaction in patchy systems is anisotropic and short-ranged. This fundamental aspect, which may or may not be complemented by a non-spherical shape~\cite{demichele_dna_patchy,Millan2015}, is at the core of the complex behaviour of these systems. Here we review some of the most common potentials used to simulate toy models of patchy particles.

In what follows we will consider systems composed of spherical particles of diameter $\sigma$. Generalisations to systems that are polydisperse in size are straightforward, although uncommon~\cite{roldan2013gelling}. By contrast, even though it is typical to simulate systems composed of identical particles, sometimes binary mixtures or even variability in the patch position, size or strength need to be taken into account~\cite{bianchi_disordered,rovigatti2013computing,romano_russo_polydispersity}. Therefore, for the sake of generality we consider the surface of each particle $i$ to be decorated with $M_i$ patches. The position of patch $\alpha$ relative to the centre of mass of particle $i$, $\mathbf{r}_i$, is specified by the vector $\mathbf{r}_{\alpha, i}$. Let $\mathbf{r}_{ij} = \mathbf{r}_j - \mathbf{r}_i$ be the distance between particle $i$ and particle $j$ and $\mathbf{r}_{ij,\alpha\beta} = \mathbf{r}_{ij} + \mathbf{r}_{\beta, j} - \mathbf{r}_{\alpha, i}$ be the distance between patch $\alpha$ on $i$ and patch $\beta$ on $j$. Furthermore, let $\lbrace \mathbf{r}_{\alpha, i} \rbrace$ be the set of vectors identifying all $M_i$ patches of particle $i$. It is common to consider patchy particles that interact through an isotropic repulsive potential, that accounts for the mutually excluded volume, plus an attractive patch-patch term that depends on the relative orientations. By using the notation introduced above, the total interaction energy between particles $i$ and $j$ can thus be written as

\begin{equation}
V(\mathbf{r}_{i}, \mathbf{r}_{j}, \lbrace \mathbf{r}_{\alpha, i} \rbrace, \lbrace \mathbf{r}_{\beta, j} \rbrace) = V_{\rm ex}(r_{ij}) + V_{\rm patch}(\mathbf{r}_{ij}, \lbrace \mathbf{r}_{\alpha, i} \rbrace, \lbrace \mathbf{r}_{\beta, j} \rbrace)
\end{equation}

\noindent
where non-bold symbols indicate the moduli of the respective vectors. The excluded-volume term is, most of the times, a hard-sphere interaction for MC simulations or a differentiable hard-sphere-like potential (\textit{e.g.} Weeks-Chandler-Anderson, inverse power) for MD simulations. However, sometimes screened electrostatic repulsions, modelled through Yukawa-like interactions, are also considered~\cite{yukawa_lysozyme,lisbona_dynamics}.

The $V_{\rm patch}$ term is a sum of the contributions of all pair of patches, viz:

\begin{equation}
V_{\rm patch} = \sum_{\alpha = 1}^{M_i} \sum_{\beta = 1}^{M_j} V_{\rm pp}(\mathbf{r}_{ij}, \mathbf{r}_{\alpha, i}, \mathbf{r}_{\beta, j}).
\end{equation}

It is common to employ attractive potentials that are, or resemble, square-well potentials (discontinuous or continuous, depending on the chosen simulation technique) and hence are functions of the patch-patch separation only, \textit{e.g.} $V_{\rm pp} = V_{\rm pp}(r_{ij,\alpha\beta})$. This family of potentials, usually called sticky-spot or point-patch potentials, have essentially two parameters: the range of the interaction $\delta$ and its strength, $\epsilon$. An example of a continuous square-well-like attraction is~\cite{russo_kf_continuous}

\begin{equation}
\label{eq:square_well}
V_{\rm pp}(r_{ij,\alpha\beta}) = -\epsilon_{\alpha\beta}\exp\left[ -\frac{1}{2} \left( \frac{r_{ij,\alpha\beta}}{\delta_{\alpha \beta}}\right)^{10} \right].
\end{equation}

A simple potential like the one of Eq.~\eqref{eq:square_well} contains the basic physics of patchy interactions: it confers a strong anisotropy and a well-defined valence to the particles. It has been used to look at the thermodynamics and dynamics of self-assembled disordered phases~\cite{russo_kf_continuous,sandalo_rovigatti,lisbona_dynamics,douglas_soft_matter}. However, as shown in Figure~\ref{fig:snapshots}(a), patch-patch radial potentials have a spherical shape whose size is fixed by the single parameter $\delta$, which therefore fully controls the bonding volume (or extent) of the patches. Consequently, the radial and angular flexibilities of a bonded pair cannot be tuned independently. As it turns out, these two quantities contribute differently to the entropy of disordered and ordered phases, going as far as determining their relative thermodynamic stability~\cite{romano2011crystallization,frank_nat_phys}. Hence, in some cases it is important to be able to separately control the angular and radial flexibilities.

A well-known potential that provides control on the patch shape is the so-called Kern-Frenkel (KF) interaction~\cite{kern_frenkel}. It features two independent geometrical parameters, $\delta$ and $\theta^{\rm max}$, that set the radial and angular width, respectively. In the KF model patches can be visualized as spherical cones (see Figure~\ref{fig:snapshots}(b)), with the patch-patch interaction being given by

\begin{equation}
V_{\rm pp}(\mathbf{r}_{ij}, \mathbf{\hat{r}}_{\alpha, i}, \mathbf{\hat{r}}_{\beta, j}) = V_{\rm SW}(r_{ij}) f(\mathbf{r}_{ij}, \mathbf{\hat{r}}_{\alpha,i}, \mathbf{\hat{r}}_{\beta, j})
\end{equation}

\noindent
where  $V_{\rm SW}$ is an isotropic square-well term of range $\sigma + \delta_{\alpha\beta}$ and depth $\epsilon_{\alpha\beta}$, the hat symbol $\hat{ }$ indicate unit vectors and $f$ is the orientation-dependent modulation term that takes the form

\begin{equation}
\label{eq:KF}
f(\mathbf{r}_{ij}, \mathbf{\hat{r}}_{\alpha, i}, \mathbf{\hat{r}}_{\beta, j}) = \left\{ 
\begin{array}{rl}  
1 & \mathrm{if} \; \begin{array}{rl} 
\mathbf{\hat{r}}_{ij} \cdot \mathbf{\hat{r}}_{\alpha, i} > \cos{\theta^\mathrm{max}_{\alpha\beta}}\\
\mathbf{\hat{r}}_{ji} \cdot \mathbf{\hat{r}}_{\beta, j} > \cos{\theta^\mathrm{max}_{\alpha\beta}}
\end{array}\\
0 & \mathrm{otherwise.}
\end{array} \right.
\end{equation}

\noindent
During the course of the years, this potential has been used to investigate the phase behaviour of many types and blends of patchy and patchy-like systems~\cite{kern_frenkel,Zhang2005,yukawa_lysozyme, rovigatti2013computing,frank_nat_phys}.

We must note that having two independent parameters that model the extent of a patch-patch interaction arises naturally when dealing with hydrophobic surfaces, and indeed the KF model has been used to reproduce quantitatively the phase diagram and crystallization process of a quasi-2D system of hard particles with hydrophobic patches~\cite{romanosm2011}. The reason is that two very short-ranged hydrophobic curved surfaces always have one contact point, which provides some negative energy. Where this contact happens on the two surfaces does not matter, and the KF model describes this kind of interaction. On the other hand,  the point-patch model might seem more natural when dealing with particles that can experience a binding interaction depending on the position of the centre of the patch: one example is if the patches are realised through a single DNA strand that has some maximum length that it can reach given its attachment point. A key difference in between the two models, that must be taken into account when considering the binding properties of two particles, is the fact that point-like models do not allow to freely rotate a particle around the vectors joining two bound particles, while the KF model does. 

In order to exploit the advantages of the KF potential in MD simulations, continuous versions of the KF interaction have been developed. These are built by taking a differentiable isotropic (often square-well-like) function and by modulating it with functions that, akin to $f(\mathbf{r}_{ij}, \mathbf{\hat{r}}_{\alpha, i}, \mathbf{\hat{r}}_{\beta, j})$ in Eq.~\eqref{eq:KF}, depend on the scalar product between the particle-particle distance and the unit vectors that identify each patch. These functions are often Gaussians or generalised Gaussians, similar to the one of Eq.~\eqref{eq:square_well}. These continuous potentials have been used to look at self-assembled ordered structures in patchy systems~\cite{PhysRevE.80.021404,janus_continuous_KF,doye_diamond,mahynski_bottomup,PhysRevE.92.042309}, but they have been employed also in modelling patchy polymers~\cite{coluzza2013sequence} and even amino acids in proteins~\cite{Coluzza2014}.

We note on passing that some of the behaviour displayed by patchy systems is also shared by other models that, in one way or another, fix the maximum number of bonds that each particle can form (the so-called \textit{valence}). For instance, common potentials developed for water and silicon employ rigid bodies complemented by charges or spheres interacting through three-body interactions that limit the ideal number of favourably interacting neighbours~\cite{stillinger_weber,molinero2008water,water_models}.

\section{The single-bond-per-patch condition}

The single-bond-per-patch condition (SBPPC) is essential to control the valence. The valence, in turn, is the most important property in determining the phase behaviour of patchy particles. Most models enforce the SBPPC through a judicious choice of the model geometry and parameters. For rather simple, and thus computationally efficient, models (such as spheres decorated with patches) the SBPPC requires quite small patches, and hence very small bonding volumes. For instance, the single geometrical parameter of the point-patch potential, the patch-patch attraction range, has to be smaller than $0.119 \sigma$ ($\sigma$ being the particle diameter) to enforce the SBPPC. By contrast, the two parameters of the Kern-Frenkel potential, the range $\delta$ and the angular width $\theta_{\rm max}$, have to fulfil the inequality $\sin(\theta_{\rm max}) \leq \left( 2 (1 + \delta\sigma)\right)^{-1}$. Under these constraints, the largest bonding volumes attainable are of the order of $\approx 10^{-3}\sigma^3$. Such small values are problematic from a technical and a fundamental point of view. Indeed, from a computational standpoint small bonding volumes make it very hard to equilibrate at low $T$. From a modelling perspective, instead, some very interesting phenomena (such as the appearance of disordered ground states~\cite{frank_nat_phys} or of reentrant gas-liquid phase separations~\cite{russo_reentrant}) require or are accessible with large bonding volumes only. Furthermore, some ``soft'' building blocks, such as telechelic star polymers or DNA nanostars, possess an intrinsically large bonding volume due to their inner flexibility~\cite{patchy_review}. The question that arises is whether it is possible to take into account such a high degree of internal flexibility in a straightforward way by devising models that are nearly as simple as the ones described above.

The issue regarding the precise value of the bonding volume can be worked around in Monte Carlo or event-driven simulations, where non-continuous constraints are easy to implement, by putting an explicit limit to the number of bonds per particle~\cite{speedy_nmax,nmax}. Operatively, this is done by adding an attractive term to the interaction between two particles only if both have fewer than the maximum number of allowed bonds.
A drawback of this approach is that the resulting model can be hardly seen as realistic and, importantly, is not amenable to be investigated \textit{via} MD simulations.
	
A more sophisticated approach has been introduced in Ref.~\cite{Sciortino2017} and employed to readily generate fully-bonded disordered configurations to model microgel particles~\cite{gnan2017silico}, and to investigate the dynamics of a simple model of vitrimers at large length- and time-scales~\cite{rovigatti_vitrimers}. The idea behind the method is to add a short-ranged three-body repulsive term to triplets of close patches in order to compensate the additional (negative) energy contribution due to the formation of extra bonds. The approach is compatible with continuous potentials, meaning that it can be used in molecular dynamics simulations. The model has a parameter, $\lambda$, that controls the energetic penalty that non-bonded patches have to pay in order to get close to a bonded pair. The model thus not only enforces the SBPPC but also introduces a bond-swapping mechanism that facilitates the dynamics at any $T$.

A cartoon depicting a bond-swapping process is presented in Figure~\ref{fig:snapshots}(c). Taking three patches $A$, $B$, and $C$ as an example, if $\lambda = 1$, then the energy of configurations where $A$ is bonded to both $B$ and $C$ (panel (ii)) or just to one of them (panels (i) and (iii)) is the same. Under this condition, the process through which the $A-B$ bond breaks and the $A-C$ bond forms is carried out at constant energy: temperature plays no role and the system can rearrange its bonding pattern even in the limit $T \rightarrow 0$. By contrast, if $\lambda > 1$, the bond swapping process becomes energetically expensive and hence thermally activated. 

\section{Monte Carlo moves}

Rototranslations and particle insertions/deletions, which are briefly introduced in the paragraphs below, are the simplest moves that allow to sample the canonical and grand canonical ensembles, respectively, for any system. However, self-assembly processes in patchy systems often occur when the bonding energy $\epsilon$ is substantially larger than the thermal energy $k_B T$ (see \textit{e.g.} Refs.~\cite{Zhang2005,doye_diamond,reentrant_gels}). For instance, it is common to simulate patchy systems with $\epsilon/k_BT \gtrsim 5$. Under these conditions, most of the patches are involved in bonds, and the Metropolis acceptance of unbiased rototranslations that attempt to break a single bond, $\exp{(-\beta \Delta E)}$, plummets from $\approx 7 \times 10^{-3}$ down to $\approx 4.5\times 10^{-5}$ as the bonding strength relative to the thermal energy increases from $\epsilon/k_BT = 5$ to $\epsilon/k_BT = 10$. As a result, the great majority of the moves will very likely be rejected, thereby greatly reducing the efficiency of the simulation. This issue can be overcome, or at least improved upon, by employing \textit{biased} Monte Carlo moves~\cite{frenkelsmit}. After a brief introduction of the most basic MC moves, we will discuss two biased MC moves that are not system-specific and can greatly enhance the effective exploration of the phase space.

\subsection{Rototranslations}

When dealing with anisotropic particles, a good sampling of the phase space requires a correct handling of both rotations and translations. In Monte Carlo simulations, the simplest moves that ensure equilibrium are pure translations and pure rotations or, as done in the PP code, rototranslations, which are combinations of the two. For both translations and rotations it is common to set the maximum (radial or angular) displacement so as to have an acceptance ratio of about 0.2~--~0.4, with the optimal value being dependent on the specific system investigated~\cite{mountain_1994,frenkelsmit}. Note that, in force of the short-range nature of the interaction potentials  (see Section~\ref{sec:potentials}), these maximum displacements will necessarily be small.

A rototranslation is carried out by choosing a random particle $i$, random angular and radial displacements and a random axis of rotation. Particle $i$ is then moved and rotated according to the values extracted. The difference in energy due to the move is then used to compute the Boltzmann factor which, in turn, is connected to the acceptance probability, as per the standard Metropolis Monte Carlo~\cite{frenkelsmit}. The only choice that needs to be made is how to handle the rotational degrees of freedom. Quaternions and explicit orientation matrices are both common choices~\cite{rapaport,allen_tildesley}. Note that floating-point arithmetic tends to disrupt the orthonormality of the data structures that store the orientational degrees of freedom. Care has to be taken to ensure that the quality of the orthonormality does not degrade too much over the course of the simulation. The \texttt{MC\_move\_rototranslate} function of the \texttt{MC.c} file shows how rototranslations are implemented in the PP code, which employs rotation matrices that are re-orthonormalised through a Gram-Schmidt procedure every time that the energy is printed (see the \texttt{MC\_check\_energy} function).

Compared to MC, rotations are a more delicate matter in MD simulations. Indeed, when dealing with rigid bodies one has to consider that particle-particle interactions exert not only central forces but also torques. In many cases these torques have complicated expressions. However, there exist guidelines that, depending on the specific type of interaction, can greatly simplify the deriving procedure~\cite{allen2006expressions}. The algorithm used to integrate the equations of motion can be also carried out in different fashions, depending on whether quaternions, rotation matrices or constraint algorithms are employed~\cite{CICCOTTI1986346,fast_shake,lincs,rapaport,rotation_matrices} and on the type of dynamics~\cite{brady_stokesian,ruslan_langevin,ouldridge_rotations}. A drawback of the most common patchy interactions (see Section~\ref{sec:potentials}) is their steepness, which negatively affects numerical stability and makes it hard (or even impossible) to use single-precision floating point arithmetic (which some GPU-powered packages offer to improve performance~\cite{lammps_gpu,gromacs_gpu}). For the same reason, extra care has to be taken when choosing the integration time step.

\subsection{Grand Canonical ensemble}

In the grand canonical ensemble (GCE) the system is in equilibrium not only with a thermal bath but also with a reservoir of particles~\cite{frenkelsmit}. Therefore, the three thermodynamic quantities that are kept fixed are volume $V$, temperature $T$ and chemical potential $\mu$. As a result, the overall number of particles $N$ fluctuates. The GCE is commonly used in Monte Carlo simulations to investigate the phase behaviour of disordered systems~\cite{wilding_gce_review}, and can be readily extended to many-component mixtures~\cite{panagiotopoulos_mixtures,horbach_mixtures}. In addition, GCE simulations can be combined with more sophisticated techniques, such as successive umbrella sampling (see Section~\ref{subsec:sus}) or Wang-Landau sampling~\cite{wang_landau} to overcome free-energy barriers. 

In GC simulations there are two additional trial moves that attempt to either add or remove a particle.
Note that for the particle addition move, the new particle must be added in a random position with a randomly distributed orientation. The two MC acceptance probabilities read

\begin{eqnarray}
{\rm acc}(N \to N + 1) & = & \min\left(1, \frac{Vz}{\Lambda^3(N + 1)}e^{-\beta \Delta E}\right)\\
{\rm acc}(N \to N - 1) & = & \min\left(1, \frac{\Lambda^3 N}{Vz}e^{-\beta \Delta E}\right)
\label{eq:gc}
\end{eqnarray}
\noindent
where $\Lambda$ is the thermal wavelength, $\beta = 1 / k_B T$, $z = e^{\beta \mu}$ is the activity and $\Delta E = E_{\rm new} - E_{\rm old}$ is the energy difference between the final and initial configurations. The implementation of the main core of the GCMC algorithm can be found in the PP code in the \texttt{MC\_add\_remove} function of the \texttt{MC.c} file.

\subsection{Aggregation-Volume-Bias moves}

Self assembly processes often require low density and temperature, specially in patchy systems. The sampling of the phase space under these conditions can be dramatically improved with the so-called aggregation-volume-bias (AVB) Monte Carlo moves~\cite{avb1,avb2}.

The AVB scheme provides two different basic moves. Detailed balance requires that both types of moves are used during the course of the simulation. The first one (hereafter referred to as the AVB-B move) attempts to form a bond between two previously unbonded particles, while the second one (the AVB-U move) attempts to break an existing bond by separating a bonded pair. Here the notion of two particles being ``bonded'' or ``unbonded'' should fulfil a criterion that need not be necessarily related to the specific model employed. However, it is often convenient to use an operative definition for the bonding between two particles that coincides with a state of low pair energy. For instance, an optimal criterion for the KF interaction is to regard two particles as bonded when they share a patch-patch bond. Once the bonding criterion has been set, the bonding (phase-space) volume is defined as the number of microscopic configurations for which two particles are bonded, $V_{\rm AVB}$. According to this definition, $V_{\rm AVB}$ depends only on the bonding criterion and not on the macroscopic thermodynamic variables (such as $T$, $\mu$, $N$, $V$, \textit{etc.}). For instance, the expressions for the bonding volume of a KF $\alpha\beta$ bond in 3D and 2D are:

\begin{align*}
V_{\rm AVB} & = \frac{\pi \sigma^3}{3} (1 - \cos\theta^{\rm max}_{\alpha\beta})^2[(\sigma + \delta_{\alpha\beta})^3 - \sigma^3] & {\rm (3D)}\\
V_{\rm AVB} & = \frac{(\theta^{\rm max}_{\alpha\beta})^2}{\pi} [(\sigma + \delta_{\alpha\beta})^2 - \sigma^2] & {\rm (2D)}.
\end{align*}

We then define the ``outer'' (phase-space) volume $V_{\rm O}$ as the number of configurations for which two particles are not bonded. It follows from its definition that $V_{\rm O} = 4 \pi V - V_{\rm AVB}$, where $V$ is the volume of the simulation box and $4 \pi$ comes from the rotational degrees of freedom. These two volumes will be used in the following to bias the acceptance of the two moves.

Finally, let $N_i$ be the number of particles that are bonded to particle $i$. The recipe for the AVB-B move is then:

\begin{enumerate}
\item Randomly select a particle $i$.
\item Randomly select a particle $j$ which is neither $i$ nor one of its bonded neighbours.
\item Move particle $j$ inside the bonding volume of particle $i$.
\item Accept the move with probability
$$\min\left( 1, \frac{(N - N_i - 1) V_{\rm AVB}}{(N_i + 1)V_{\rm O}} \exp(-\beta \Delta E)\right).$$
\end{enumerate}

Note that the insertion of particle $j$ in the bonding volume of particle $i$ must be carried out uniformly. In other words, the probability that, at the end of a trial move, $i$ and $j$ are in a specific mutual arrangement must be the same for every microscopic configuration that fulfils the bonding criterion. The PP code shows how to perform this operation for KF patches in the \texttt{place\_inside\_vbonding} function of the \texttt{utils.c} file.

In order to perform an AVB-U move one has to

\begin{enumerate}
\item Randomly select a particle $i$.
\item Randomly select a particle $j$ that is bonded with $i$. If $i$ has no neighbours (that is, if $N_i = 0$) then outright reject the move.
\item \label{itm:move} Move particle $j$ outside of the bonding volume of $i$, so that at the end of this step $i$ and $j$ are no longer bonded.
\item Accept the move with probability
$$\min\left( 1, \frac{N_i V_{\rm O}}{(N - N_i)V_{\rm AVB}} \exp(-\beta \Delta E)\right).$$
\end{enumerate}

Note that step~\ref{itm:move} can be performed by randomly inserting $j$ in the box with a random orientation till $i$ and $j$ are no longer bonded.

The ratio between the two volumes defined above, $r \equiv V_{\rm AVB} / V_{\rm O}$, is what biases the acceptance of the AVB-B move. Since the particle bonding volume is always (much) smaller than the overall volume, the bias \textit{lowers} the acceptance probability of the move. However, at low temperature this is rendered unimportant by the very large value of the Boltzmann factors associated to the creation of additional bonds.
By contrast, $1/r$, which is much larger than $1$, biases the acceptance of the AVB-U move. Therefore, the larger the value of $r$, the better AVB will perform compared to a standard MC algorithm. As a result, the AVB algorithm shines when used in conjunction with short-ranged potentials (small $V_{\rm AVB}$) and low-density systems (large $V_{\rm O}$). There is an additional reason why AVB moves perform better at low density: it dramatically increases the speed with which particles move around the simulation volume, which is a common efficiency bottleneck for low-density systems when only single-particle rototranslations are employed.

We note on passing that there exists a third type of AVB moves which takes a particle $i$ bonded with particle $j$ and insert it in the bonding volume of a third particle~\cite{avb2}. It can greatly enhance the sampling in specific cases such as chain-forming systems~\cite{russo_reentrant}.

\begin{figure}[t]
\includegraphics[width=0.45\textwidth]{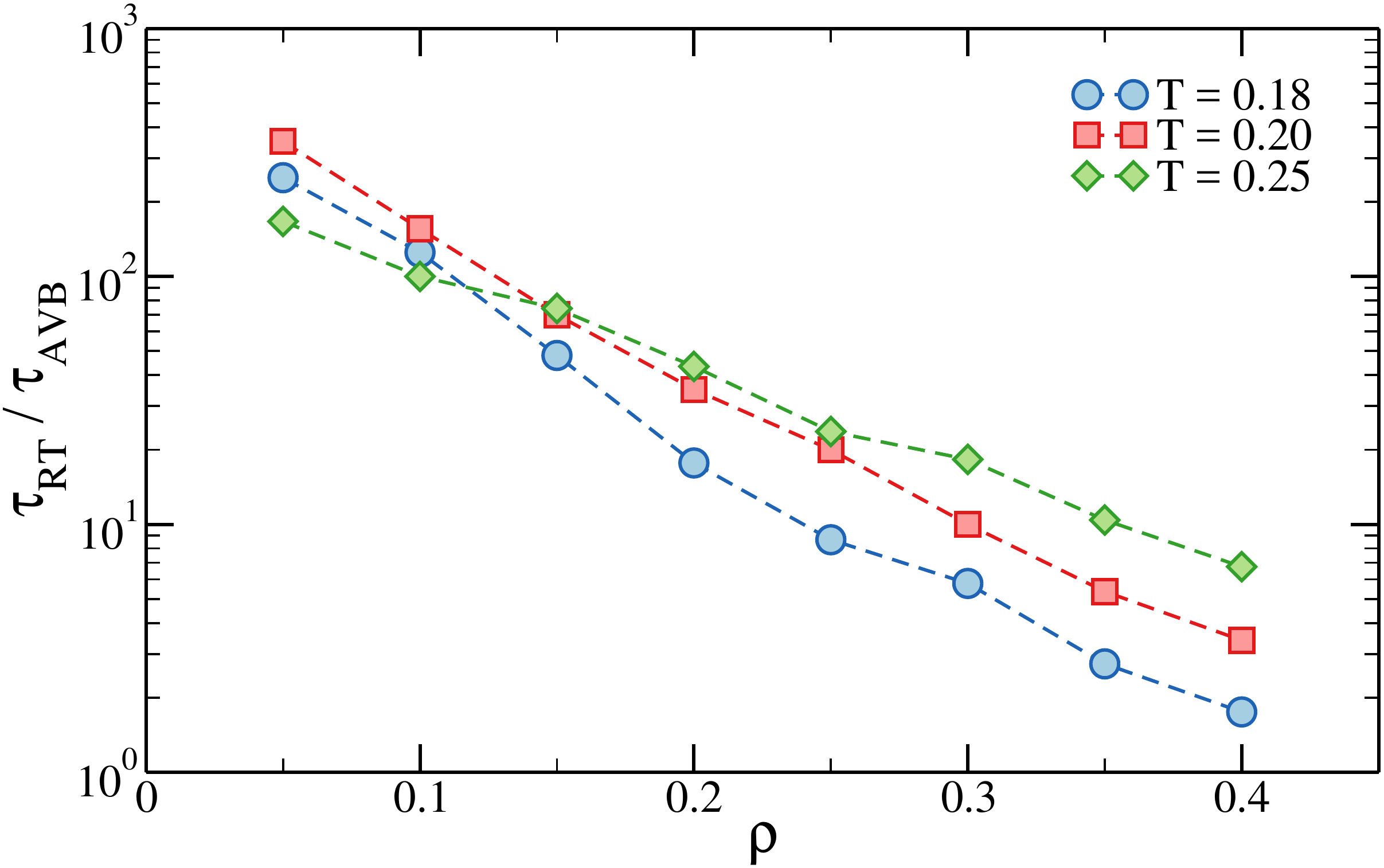}
\caption{\label{fig:comparison_avb}The ratio between the autocorrelation time of the energy of systems simulated without and with AVB moves as a function of density for different $T$. Assuming that the computational cost of the two procedures is the same (see text), this ratio represents the gain achieved when using AVB.}
\end{figure}

In order to benchmark the performance of the AVB scheme we simulate several systems made of $N = 500$ tetravalent patchy particles at different $\rho$ and $T$. For each one we compute the autocorrelation function of the energy. We then estimate the time $\tau$ at which the autocorrelation function takes the value $1/e$ and use it as a proxy for the simulation efficiency. Figure~\ref{fig:comparison_avb} shows the ratio between the autocorrelation time for systems where AVB moves were not attempted, $\tau_{\rm RT}$, and attempted $50\%$ of the time, $\tau_{\rm AVB}$. As mentioned above, the AVB scheme performs better at low density. For very small values of the density ($\rho \leq 0.1$), which are nonetheless still relevant to many self-assembly processes~\cite{russo_kf_continuous}, AVB enhances the efficiency of the simulation by more than two orders of magnitude. By contrast, the dependence on temperature is less dramatic, albeit non-monotonic, and more system-dependent.

A basic implementation of the AVB algorithm is contained in the \texttt{avb.c} and \texttt{avb.h} files of the PP code. We note that the PP code runs roughly $20\%$ -- $40\%$ slower when AVB moves are enabled. This is merely an implementation issue and mostly due to the lack of optimisation that stems from the educational nature of the PP code. With a properly optimised code the AVB scheme is only marginally slower than a regular Monte Carlo simulation.

We conclude this section by mentioning that AVB moves and GCMC simulations complement each other very well. Indeed, when simulating low-temperature systems the most probable GC deletions are those attempted on monomers. Unfortunately, under normal conditions it is most likely to attempt to remove a monomer generated by a previous GC addition rather than a monomer that spontaneously broke off all of its bonds. As a result, many GC moves are essentially wasted. However, when the AVB scheme is applied the rate of monomer formation/depletion is greatly enhanced by the AVB-U and AVB-B moves, respectively, and thus the number of effective GC moves increases.

\subsection{Virtual Move Monte Carlo}

The Virtual Move Monte Carlo (VMMC) is a Monte Carlo cluster-move algorithm proposed originally by Whitelam and Geissler~\cite{vmmc1} and designed to improve relaxation times in strongly-interacting, low-density systems and to better approximate diffusive dynamics in these systems. As it often happens, the algorithm has proved to be useful in many other contexts, such as polymeric systems~\cite{usevmmc1,sulc2012}. The VMMC cluster move, quite complicated to implement in its original version, has been presented in a more simple way by the same authors~\cite{vmmc2}, and more recently an easier version to program has been published by R\r{u}\u{z}i\u{c}ka and Allen~\cite{vmmc3}. The idea behind the algorithm is to avoid one of the shortcomings of standard cluster algorithms, whereby the clusters are built based upon the system microstate \emph{before} the move, and any move that attempts to merge or break clusters has to be rejected in order to satisfy detailed balance. This means that cluster moves need to be mixed with standard single-particle moves, and although they do speed up the diffusion of clusters, the pattern of the interactions is changed mostly by single-particle moves. In the context of patchy particles, where patch-patch bonds are the natural criterion to define clusters, cluster moves based only on the system microstate before the move will never change the bonding pattern by construction. To overcome this limitation, Whitelam and Geissler have designed the VMMC cluster move, a ``virtual move'' that builds a cluster based upon the \emph{move} as well as the starting microstate: particles are recruited into the cluster depending on how much their energy changes with the move, rather than how strongly they interact with their neighbours. Essentially, the cluster is chosen on the fly by starting to move a particle and then recruiting all those neighbouring particles that would rather move than remain still and pay an energetic cost.

Here we describe the most recent iteration of the algorithm proposed in Ref.~\cite{vmmc3}, implemented in the function \texttt{VMMC\_dynamics} of the \texttt{VMMC.c} file of the PP code.

\begin{enumerate}
\item\label{al:start} Choose a move (a random displacement for translations, an axis and an angle for rotations).
\item Choose a seed particle for the move, and add it to the cluster. The cluster will contain all those particles that will be moved.
\item\label{al:list} Build or update a list $\mathfrak{L}$ of particle pairs that will have their pair energy changed if the cluster is moved. Note that, by construction, one and only one particle in each pair belongs to the cluster, since the cluster moves as a single object and hence the pair interaction energy of particles that belong to the cluster will not be affected by the move. Also note that $\mathfrak{L}$ will contain particle pairs that are affected from the cluster moving both away and towards them.
\item Choose a random pair $l=(i,j)$ in $\mathfrak{L}$. Call $i$ the particle in the pair $l$ that belongs to the cluster and $j$ the particle that does not. Remove this pair from the list.
\item Compute (or recover from bookkeeping) $u(i,j)$, the energy of the pair in the starting microstate.
\item Compute $u(i',j)$, the energy of the pair after $i$ has been moved and $j$ has not.
\item Compute $u(i,j')$, the energy of the pair after $j$ has been moved and $i$ has not.
\item\label{al:rec1} Extract a random number $r \in \left[ 0, 1 \right]$. If $\exp\{{-\beta(u(i',j)-u(i,j))}\} > r$, insert $j$ in a list $\mathfrak{C}$ of particles that are candidates for recruitment in the cluster.
\item\label{al:rec2} Extract a new random number $q \in \left[ 0, 1 \right]$. If $\exp\{-\beta(u(i,j')-u(i,j))\} > q$, recruit $j$ into the cluster and remove it from $\mathfrak{C}$.
\item If $\mathfrak{L}$ is not empty, go to step \ref{al:list}. If $\mathfrak{L}$ is empty, the cluster has been built. Please note that the cluster may contain just the seed particle.
\item\label{al:end} If $\mathfrak{C}$ is not empty, reject the move. This step is necessary to enforce balance~\cite{vmmc1}. Otherwise, accept the cluster move.
\end{enumerate}

We point the reader to Ref.~\cite{vmmc1} and Ref.~\cite{vmmc3} for a detailed discussion of how this algorithm is derived. Here we will just elaborate on some of the less obvious points of the algorithm. For example, the random choice of an element in $\mathfrak{L}$, rather than processing the pairs as the algorithm finds them, is needed to ensure super-detailed balance, \textit{i.e.}, that the algorithm is in principle able to perform the exact reverse move immediately. Also, the fact that $\mathfrak{C}$ needs to be empty for the move to be accepted is done for the same reason. The quantities in steps~\ref{al:rec1} and ~\ref{al:rec2} are chosen in such a way to build a cluster move that will be automatically accepted in step~\ref{al:end} if $\mathfrak{C}$ is empty.

Although significantly more difficult to implement than a standard single-particle or cluster move, this algorithm has a number of benefits that make its use very beneficial in a variety of situations. The fact that it is the move that dictates the cluster allows for internal relaxation of strongly-interacting regions: particles are recruited into the cluster only if not doing so would increase the energy (on average). Strongly interacting pairs are not necessarily recruited together: in the extreme case of $u(i,j)$ very negative, if $u(i',j)$ is also very negative (the relative move does not affect the energy) particle $j$ is \emph{not} recruited into the cluster (see step \ref{al:rec1}). This allows, for example, for relative rotation of clusters or for naturally arising pivot and crankshaft moves in polymeric systems.

In our experience, we have found that VMMC is a good algorithm for self-assembling and polymeric systems due precisely to its ability of finding the natural moves that help relaxing the system. On the other hand, the algorithm has a few drawbacks. Firstly, care must be taken in the choice of the move at point 1 above: a large move will easily recruit the whole system into the cluster, essentially wasting computer time by building a large cluster for a move that, even if accepted, is effectively useless. This would not happen frequently in the low-density scenario for which the algorithm was designed, but nevertheless it should be taken into account. Secondly, the algorithm is not straightforward to program in an efficient way.

In terms of performances, the original paper~\cite{vmmc1} highlights that the VMMC algorithm is best suited for low-density, low-temperature systems, where the system is largely composed of clusters of particles. In particular, VMMC greatly speeds up the equilibration of inter-cluster interactions essentially by promoting a much faster diffusion of clusters than single-particle MC moves. Importantly, clusters can merge and break while moving, which is forbidden in standard cluster moves. Since each virtual move is much more computationally expensive than a standard single-particle move, the benefits of VMMC are worth the extra effort only when the diffusion of clusters is the bottleneck of the simulation.

\section{Free-energy methods}

The level of control and tunability of patchy models has opened the way to a deeper understanding of the phase behaviour of complex systems with directional interactions~\cite{zhang2004self,glotzer2007anisotropy,sciortino2016basic,patchy_review}. For the disordered phases (liquid and gas), patchy models were developed to study both theoretically and computationally the process of self-assembly, with particular emphasis on the polymerization~\cite{sciortino2007self} and gel transitions~\cite{russo_kf_continuous}. For ordered (crystal) phases, patchy particles have been shown to be able to describe the crystallization of open crystalline structures~\cite{chen2011directed,saika2011nucleation,romano2012phase}, most notably the diamond crystal for 4-patch models~\cite{doye_diamond,romano2011crystallization,romano_russo_polydispersity}. Self-assembly can have a deep impact on the phase behaviour of the system, and patchy particles were the ideal platform where new transitions and new types of phase diagrams were discovered. Empty liquids~\cite{bianchi2006phase}, re-entrant and topological phase transitions~\cite{russo_reentrant}, amorphous ground states~\cite{frank_nat_phys}, are all examples where new types of phase behaviour were discovered in the context of patchy particles.

Most notably these discoveries were made possible by the use of highly-efficient free-energy methods. We should note that free-energy methods can be divided into two broad categories: i) the ones that attempt to recover the free energy of a thermodynamic phase, and ii) the ones that aim at finding the \emph{potential of mean force}, i.e. the projection of the free energy along a \textit{reaction coordinate} or \textit{order parameter}. In the next sections we will briefly discuss the former  and then describe more in depth one of the latter.

\subsection{Direct evaluation}\label{sec:direct}

A phase diagram of a substance shows under which thermodynamic conditions a phase (or a set of phases) is in equilibrium. 
From a simulation standpoint, there exist several methods to straightforwardly estimate the phase boundaries between disordered phases (one of which will be thoroughly discussed in the next section)~\cite{landau2014guide,virnau2004calculation}. By contrast, the presence of ordered phases greatly complicates the numerical evaluation of the phase behaviour. Quite remarkably, the main challenge is not posed by the actual calculation of the relative free energy between the phases (which, in turn, dictates their relative stability), but by the fact that, in general, it is not possible to know \textit{a priori} what these phases are in terms of symmetry, unit cells, \textit{etc}. Indeed, while for simple particles such as hard spheres an educated guess does often suffice, more complicated building blocks such as patchy or non-spherical particles might require approaches that go beyond ``hand''-based or brute-force procedures. An efficient way of searching the vast space of possible lattice candidates is provided by the so-called floppy box method. The basic idea is to simulate a small number of particles in a box of variable shape in the isothermal-isobaric ensemble~\cite{PhysRevLett.103.188302}, and collect the most stable structures in the range of temperatures and pressures of interest. This method has been applied to, among others, non-convex objects~\cite{PhysRevLett.107.155501,doi:10.1021/nl100783g} and one-patch particles~\cite{vissers2013predicting}. In parallel, the idea of leveraging genetic algorithms to efficiently compile a list of candidate structures has surfaced, and it has been employed in patchy systems to find those ordered arrangements that are most favourable from an energetic standpoint~\cite{genetic_2D,genetic_3D}. The two methods tend to perform similarly, and the best choice depends on the system under investigation~\cite{genetic_vs_floppy}.

Once the most promising candidate structures have been singled out, their relative stability is evaluated by computing the free-energy difference with respect to a system of known free energy. This is usually done by employing a combination of Hamiltonian and thermodynamic integration techniques. Well-known methods that build upon this idea are the Frenkel-Ladd~\cite{fladd} and Einstein Molecule~\cite{vegamolecule} methods. A more thorough discussion on this topic falls out of the scope of the present review, and we refer the interested reader to Ref.~\cite{vega2008determination}.

\subsection{Potential of mean force}
Another approach is to reconstruct the free-energy landscape by sampling rare fluctuations of one or more relevant order parameters. The last decade has seen a flourishing of such methods, as for example Wang-Landau sampling~\cite{landau2014guide} and Metadynamics~\cite{laio2008metadynamics,barducci2008well}. Here we focus on methods based instead on the Umbrella Sampling technique, which have had the biggest impact on the literature on Patchy Particles. In this section we first briefly review the basics of the Umbrella Sampling method, and then focus on a variant (Successive Umbrella Sampling) which has found widespread adoption.

\subsubsection*{Umbrella Sampling}

Umbrella sampling (US)~\cite{torrie1977nonphysical,frenkel2001understanding} is a method for enhancing the sampling of regions of configurational space which have a small weight (as measured by the Boltzmann distribution) under equilibrium conditions. By enabling the sampling of rare fluctuations, Umbrella Sampling allows one to reconstruct the free-energy profile even in regions of low statistical weight, such as on top of free-energy barriers.
This is achieved by modifying the Markov chain with an external potential term which depends on the relevant order parameter for the transition of interest. The most immediate advantage of US over its competitors is that simulations of a system with a biased potential run in equilibrium, and the bias can be removed to recover the unbiased (true) probability distribution of the order parameter.


In Umbrella Sampling, a biasing potential $\eta(x)$ is added to the Hamiltonian of the system
$$
\mathcal{H}'=\mathcal{H}+\eta(x)
$$
where $\mathcal{H}$ is the base Hamiltonian (\textit{e.g.} the patchy potential), and $x$ is the order parameter whose distribution $\mathcal{P}(x)$ we wish to measure. By running a simulation with the new Hamiltonian $\mathcal{H}'$, and measuring its order parameter distribution, $\mathcal{P}'(x)$, we can readily recover the original distribution as
\begin{equation}\label{eq:us}
\mathcal{P}(x)=\mathcal{P}'(x)\exp{\left(\beta\eta(x)\right)}
\end{equation}
By choosing an appropriate $\eta(x)$ it is thus possible to sample the order parameter for arbitrary values of $x$, and then recover the original unbiased distribution with Eq.~\eqref{eq:us}.
A standard choice is to use an harmonic potential, $\eta(x)=k(x-x_0)^2$, that forces the system to sample regions around $x=x_0$ with fluctuations that depend on the coefficient $k$.
A typical use of Umbrella Sampling to estimate the free energy landscape is to perform several biased simulations centered around adjacent values of $x_0$, such that the fluctuations of the order parameter $x$ of two consecutive simulations have a significant overlap.
The full distribution $\mathcal{P}(x)$ is then obtained by combining the order parameter histograms of different simulations together. A statistically self-consistent way to do this is \textit{via} the Weighted Histogram Analysis method~\cite{histogram_reweighting}. From knowledge of the distribution function $\mathcal{P}(x)$ we can then obtain the free energy (or potential of mean force) as
\begin{equation}\label{eq:fus}
F(x)=-k_BT\,\log{\mathcal{P}(x)}
\end{equation}

Umbrella Sampling is an equilibrium method that should give the correct free energy independently of the choice of the biasing potential. However, in practice, care has to be taken to ensure appropriate sampling of the configurational space. When dividing the order parameter domain in several intervals (each centered at a different value of $x_0$), ensuring a proper sampling within each window often requires choosing different potentials for each window. In the case of harmonic potentials, for example, different intervals often require different values of the elastic constant $k$ depending on the slope of the free energy around $x_0$. The higher the slope of the free energy, the higher the value of $k$ one needs to set in order to sample correctly the configurations around $x_0$. But since the slope of the free-energy is not known \emph{a priori}, one often has to tune the values of $k$ during the course of the simulation. To minimize this problem, two diametrically opposite strategies have been adopted.

One possibility is to sample the free energy landscape in as few intervals as possible. A way to do this is to use specific forms of the biasing potential based on theoretical expectations for the barrier. We give here an example for crystal nucleation processes. A good order parameter for crystallization processes is the size of the largest crystalline cluster~\cite{prestipino2018barrier,speckhardspheres}, which we denote here as $x$. Starting from the metastable liquid phase ($x=0$), we want to use Umbrella Sampling to access rare fluctuations, corresponding to the appearance of a crystalline nucleus of size $x$. Classical Nucleation Theory~\cite{russo2014new} predicts the free energy barrier for nucleation as
\begin{equation}
 \beta\Delta F(x)=-|\Delta\mu|x^{2/3}(x^{1/3}-1.5\,x_c^{1/3})
\end{equation}
where $\Delta\mu$ is the chemical potential difference between the two phases, $x$ is the number of particles in the largest crystal nucleus, and $x_c$ is the critical nucleus size.
In order to explore states on top of the barrier, we modify the Hamiltonian with a biasing potential that exactly cancels the barrier
\begin{equation}\label{eq:cntus}
\eta(x)=-\beta\Delta F(x)
\end{equation}
This variant, called Classical Nucleation Theory Umbrella Sampling scheme (CNT-US), has been introduced in Ref.~\cite{russo2014new}. The only parameter needed is an estimate of $\Delta\mu$. This can be obtained for example with the following relation $\Delta\mu=H_\text{fusion}(1-T/T_m)$ (where $H_\text{fusion}$ is the enthalpy of fusion, while $T_m$ is the melting temperature),
or by computing the absolute free energies of the bulk phases, as described in Section~\ref{sec:direct}.
Once a reasonable approximation of $\Delta\mu$ is obtained, several independent simulations at different values of $x_c$ are run. When $x_c$ is close to the true critical size, the corresponding simulations will exhibit large fluctuations, as the bias in Eq.~\ref{eq:cntus} is effectively cancelling the nucleation barrier. All other simulations (which exhibit small fluctuations in $x$) are quickly discarded. Instead, the simulation which maximazes the fluctuations is used to reconstruct the whole barrier by using Eqs.~\ref{eq:us} and \ref{eq:fus}.

The second approach to free energy calculations is embodied by a technique called Successive Umbrella Sampling (SUS). Here, instead of targeting a uniform sampling of a large portion of the order parameter space, the aim is to divide the order parameter space in as many small intervals as possible.

\subsubsection*{Successive Umbrella Sampling}
\label{subsec:sus}
 In Successive Umbrella Sampling~\cite{virnau2004calculation} simulations, the order parameter space is divided in many small intervals that are sampled consecutively. While the approach is general, we will present it here for the choice $x=N$, i.e. when the order parameter is simply the number $N$ of particles in the simulation. This choice is appropriate for those transitions in which the density $\rho=N/V$ is a good order parameter. To sample fluctuations in the number of particles N, each SUS simulation is performed in the grand-canonical ensemble. In the context of patchy particles, the method has been used to study liquid-gas transitions, for example in re-entrant fluids~\cite{russo_reentrant},  mixtures~\cite{rovigatti2013computing}, dipolar interactions~\cite{rovigatti2011no}, and isotropic-nematic phase boundaries~\cite{sus_demichele}.

In SUS, the order parameter $N$ is divided in many intervals (also called windows) of size $k$, such that the number of particles allowed in each interval $i$ is
$$
N\in\left\{N_i,N_{i+1},\cdots,N_{i+k-1}\right\}
$$
For each interval, a simulation is run where its number of particles is constrained by reflecting boundary conditions at $N_{i-1}$ and $N_{i+k}$. In other words, grand-canonical moves that would bring the number of particles outside the allowed range are rejected.
The size of each interval $k$ is often set equal to $k=2$ to maximize the number of intervals, and minimizing the overlap between them
$$
\left\{1,2\right\},\left\{2,3\right\},\cdots,\left\{N-2,N-1\right\},\left\{N-1,N\right\}
$$

During a simulation sampling the interval $i$, the number of times $N$ was visited is used to construct the histogram $H_i(N)$. If a move is rejected due to reflective boundary conditions, the histogram is still updated for the value of $N$. In the simplest implementation, all windows are simulated independently and without the use of biasing potentials.

In each interval (assuming without loss of generality that $k=2$) the free energy difference between $N_i$ and $N_{i+1}$ is given by
$$
F(N_{i+1})-F(N_i)=-k_BT\log\frac{H_i(N_{i+1})}{H_i(N_i)}
$$
The total free energy is obtained by combining the different intervals
$$
F(N)-F(0)=-k_BT\log\frac{\mathcal{P}(N)}{\mathcal{P}(0)}
$$
where $\mathcal P(N)$ is the unnormalised probability distribution
\begin{equation}\label{eq:sus}
\frac{\mathcal{P}(N)}{\mathcal{P}(0)}=\frac{H_N(N)}{H_N(N-1)}\frac{H_{N-1}(N-1)}{H_{N-1}(N-2)}\cdots\frac{H_1(1)}{H_1(0)}.
\end{equation}

The SUS technique provides some definitive advantages over traditional Umbrella Sampling simulations. The first one is that, since the windows are very small, even unbiased simulations are able to sample all the states within the window, without the use of auxiliary potentials. If needed, umbrella potentials can still be used, and it is often possible to extrapolate the potential in one window from the measured free-energy difference in the previous window. The second big advantage is that SUS allows to trivially (and massively) parallelize the computation of free energies, as the calculation is divided in a large number of independent windows. All this while retaining the advantage of the traditional Umbrella Sampling scheme, which is the use of equilibrium simulations that can run without any prior knowledge of the free-energy barrier.

\begin{figure}[t]
\includegraphics[width=0.45\textwidth]{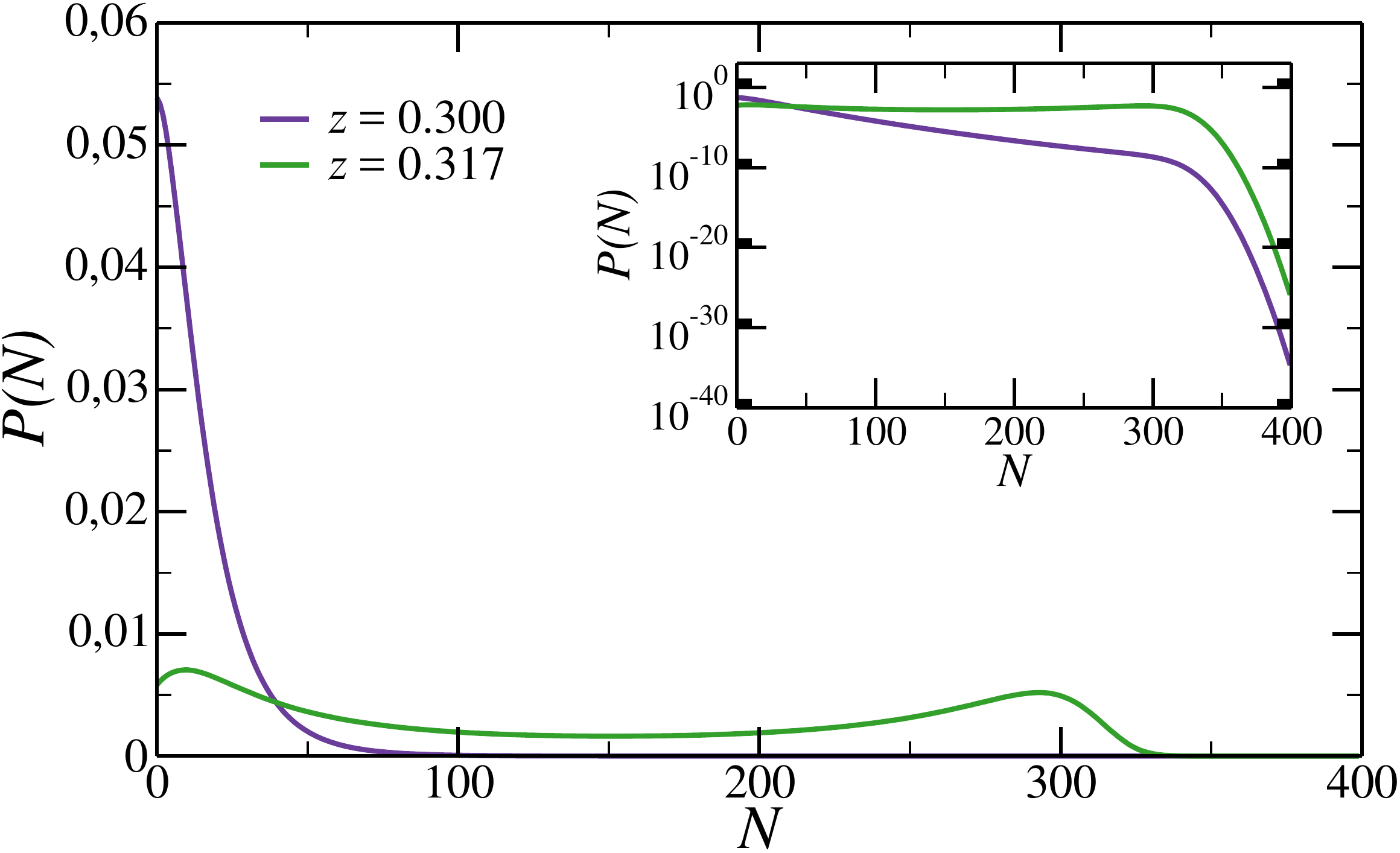}
\caption{\label{fig:sus}The probability distribution of having $N$ particles, $\mathcal P(N)$, for a system of KF tetravalent patchy particles with $\delta = 0.26$ and $\cos\theta^{\rm max} = 0.92$ simulated at $z = 0.300$ and $T = 0.160$, below the critical temperature (violet curve). Here we also show the $\mathcal P(N)$ reweighted so that the area below the peaks of the two coexisting phases is equal (green curve). The inset shows the same data on a log-lin scale to show the tens of orders of magnitude accessed.}
\end{figure}

An example of the output of a SUS simulation performed with the PP code is shown in Fig.~\ref{fig:sus}. Here the system under study is composed of tetravalent patchy particles interacting through a KF potential with parameters $\delta = 0.26$ and $\cos\theta^{\rm max} = 0.92$, simulated at $T = 0.160$, $V = 667\,\sigma^3$ and $z = 0.3$. Since the simulation has been run at a sub-critical temperature, there exist some values of $z$ for which the $\mathcal P(N)$ is double-peaked. In general, SUS simulations allow for an accurate sampling over many orders of magnitude (see inset of Figure~\ref{fig:sus}), which makes it possible to evaluate the probability distribution at different $z$ by histogram reweighting techniques~\cite{histogram_reweighting}. Here we have changed $z$ until the area below the peaks associated to the gas and liquid phases is equal (see green curve in Fig.~\ref{fig:sus}), a condition which is commonly associated to phase coexistence.

In the PP code, the histogram is updated in the \texttt{do\_SUS} function of the \texttt{MC.c} file, while the GC moves, and the logic for the rejection of the moves that attempt to break out of the window's boundaries, are implemented in the function \texttt{MC\_add\_remove} of the same file.

\section{Conclusions}

In this paper we have reviewed some of the most common models and most useful techniques that have been used to simulate systems of patchy particles, with a special focus on Monte Carlo methods.
In contrast with atomic and molecular systems, when simulating patchy particles the computationally intensive part is not the evaluation of the pair interactions or of the forces, since the potential is usually simple and very short ranged. On the the other hand, the short-range nature of the interaction sets the smallest length scale that needs to be sampled, forcing very small single-particle moves in Monte Carlo simulations and small values of the time integration steps in molecular dynamics.

A further challenge resides in the fact that the interesting behaviour in this class of systems is almost inevitably found at very low temperature, \textit{i.e.}, when a bond between two particles is of the order of 5-10 times (or more) the thermal energy. This contributes to make it challenging to achieve equilibrium without the use of advanced Monte Carlo techniques such as the ones described here.


Since it is impossible to devise a Monte Carlo move that will speed up sampling for any system and state point, we have focussed on techniques that are general enough to be useful in many use cases and showed how they perform on a typical and well known patchy particle system. Since some of the techniques treated are not straightforward to implement, we also provide an (as simple as possible) implementation in a tutorial code freely available online~\cite{pp_code}. We believe such educational tool to be a useful starting point to build up bespoke codes tailored to study systems that would not be treatable otherwise.

\section*{Acknowledgements}
All authors are greatly indebted to Francesco Sciortino, whose vision and brilliance are a constant source of inspiration. LR acknowledges support from the ERC Consolidator Grant 681597 MIMIC. JR acknowledges support from the ERC Grant DLV-759187 and the Royal Society University Research Fellowship. We thank V. Bianco, P. Handle and E. Locatelli for their insightful feedback.

\bibliography{patchy_review}

\end{document}